\documentclass[twocolumn,showpacs,preprintnumbers]{revtex4}
\usepackage{amsmath}
\usepackage[dvips]{graphicx}
\usepackage{bm}

\begin{document}

\title{Angular dependence of magnetic properties in Ni nanowire arrays}
\author{R. Lav\'{\i}n$^{1,2,3}$}
\author{J. C. Denardin$^{1,3}$}
\author{J. Escrig$^{1,3}$}
\author{D. Altbir$^{1,3}$}
\author{A. Cort\'{e}s$^4$}
\author{H. G\'{o}mez$^5$}

\affiliation{$^1$ Departamento de F\'{\i}sica, Universidad de
Santiago de Chile, USACH, Av. Ecuador 3493, Santiago, Chile\\
$^2$ Facultad de Ingenier\'{\i}a, Universidad Diego
Portales, Ej\'{e}rcito 441, Santiago, Chile\\
$^3$ Centro
para el Desarrollo de la Nanociencia y Nanotecnolog\'{\i}a,
CEDENNA\\
$^4$ Departamento de F\'{i}sica, Universidad T\'{e}cnica Federico Santa
Mar\'{i}a, Av. Espa\~{n}a 1680, Casilla 110 V, Valpara\'{i}so,
Chile\\
$^5$ Instituto de Qu\'{\i}mica, Facultad de
Ciencias, Universidad Cat\'{o}lica de Valpara\'{\i}so, Casilla
4059, Valpara\'{\i}so, Chile}


\begin{abstract}
The angular dependence of the remanence and coercivity of Ni
nanowire arrays produced inside the pores of anodic alumina
membranes has been studied. By comparing our analytical
calculations with our measurements we conclude that the
magnetization reversal in this array is driven by means of the
nucleation and propagation of a transverse wall. A simple model
based on an adapted Stoner-Wohlfarth model is used to explain the
angular dependence of the coercivity.
\end{abstract}

\maketitle

\section{Introduction}

Arrays of magnetic nanowires have attracted considerable interest
due to their promising technological applications \cite{KDA+98,
CKA+99, WAB+01, GBH+02} mainly in high-density information
storage. Electrodeposition in polycarbonate and alumina membranes
is the preferred technique used in the fabrication of these
systems \cite{MFS+08, MF95}, leading to the production of highly
ordered arrays of magnetic nanowires inside the pores of membranes
\cite{MF95, NWB+01, WLZ+06, EAJ+07, LLI+07}. Different groups have
investigated the role of magnetostatic interactions and the
influence of the size of nanowires on the magnetic properties of
these systems \cite{ELP+08, Hertel02, VPH+04, LDE+08}. Recently
Escrig \textit{et al} \cite{ELP+08} investigated the dependence of
the coercivity of Ni nanowire arrays when the external magnetic
field is applied parallel to the nanowire axis. However, for
applications in devices that use longitudinal/perpendicular
magnetic recording, it is important to investigate the angular
dependence of the magnetization. It is expected that the direction
of the external magnetic field with respect to the wire axis
strongly influences the magnetic properties of the array.

The properties of virtually all magnetic materials are controlled by domains
separated by domain walls. Measurements on elongated magnetic nanostructures%
\cite{WDM+96} highlighted the importance of nucleation and propagation of a
domain wall between opposing magnetic domains in the magnetization reversal
process. Three main idealized modes of magnetization reversal have been
identified  and occur depending on the geometry of the particles \cite%
{Hertel02, FSS+02, LAE+07}. These mechanisms are known as \textit{coherent
rotation}, C, with all the spins rotating simultaneously; \textit{vortex wall%
}, V, in which spins invert progressively via propagation of a vortex domain
wall; and \textit{transverse wall}, T, in which spins invert progressively
via propagation of a transverse domain wall. If a wire is thin enough, the
exchange interaction forces the magnetization to be homogeneous through any
radial cross section of the particle \cite{HK04}. In this way a transverse
wall is the preferred mode in thin ferromagnetic wires (diameters $d$ $<$ 60
nm).

To investigate the angular dependence of the reversal of the
magnetization in nanoparticles, several models have been proposed.
In particular, the Stoner-Wohlfarth model has been used to
calculate the angular dependence of the coercivity when the
reversal of the magnetization is driven by coherent rotation
\cite{SW48}. However, Landeros \textit{et al} \cite{LAE+07} have
shown that coherent rotation is present only in very short
particles, namely, when the length of the particles is similar to
the domain wall width. On the other side, Aharoni \cite{Aharoni97}
calculated the angular dependence of the nucleation field in an
ellipsoid
when the reversal is driven by curling rotation. Escrig \textit{et al} \cite%
{EDL+07} extended this expression for magnetic nanotubes. Finally,
Allende \textit{et al} \cite{AEA+08} presented an expression for
the angular dependence of the coercivity in magnetic nanotubes
when the reversal of the magnetization is driven by a transverse
domain wall. However, the angular dependence of the coercivity in
magnetic nanowire arrays has not been modeled yet. Thus, usually
experimental results for the angular dependence
of the coercivity are interpreted using the Stoner-Wohlfarth model \cite%
{SW48}.

In this paper we present an analytical model that allows us to
investigate the angular dependence of the coercivity and remanence
for Ni nanowire arrays, considering the different modes that can
be present, as a function of wire geometry. Additionally,
experimental data for the switching field of high-aspect ratio Ni
nanowires will be compared with this analytical model.

\section{Experimental Methods}

Our approach to the preparation of magnetic Ni nanowires arranged
in hexagonally ordered, parallel arrays is based on the
combination of two complementary aspects, namely, (i) the use of
self-ordered anodic aluminum oxide (AAO) as a porous template and
(ii) the electrodeposition of Ni in the cylindrical pores.

\textit{Anodic aluminum oxide }is obtained from the
electrochemical oxidation of aluminum metal under high voltage
(usually 20-200 V) in aqueous acidic solutions \cite{MF95}. Under
certain proper sets of experimental conditions (acid nature and
concentration, temperature, and applied voltage), the
electrochemically generated layer of alumina displays a
self-ordered porous structure. Cylindrical pores of homogeneous
diameter are thus obtained, with their long axis perpendicular to
the plane of the alumina layer and ordered in a close-packed
hexagonal arrangement. With our method \cite{FRG+05}, first and
second anodizations of Al have been carried out in 0.3\textit{M}
oxalic acid under 40 V at 20
%TCIMACRO{\U{ba}}%
%BeginExpansion
${{}^o}$%
%EndExpansion
C. Subsequently, the pores were widened in a 0.085\textit{M} H$_{3}$PO$_{4}$
solution at 37
%TCIMACRO{\U{ba}}%
%BeginExpansion
${{}^o}$%
%EndExpansion
C, obtaining pores of $d=2R=50$ nm diameter and a center-to-center distance of $%
D=100$ nm.

\textit{Electrodeposition} was performed at a constant potential
(dc electrodeposition at -1.0 V). To facilitate the electric
contact, a very thin Au-Pd layer was sputtered on one side of the
membrane, followed by the electrodeposition \cite{ELP+08, CRP+09}
of a thicker nickel layer to achieve full pore sealing. The
potentiostatic condition allows us to have more precise control of
the electrochemical reaction, and then more accurate control of
the growth of the Ni nanowires \cite{ELP+08}.

The morphology and the single-crystal structure of the individual
nanowires after dissolution of the template were subsequently
investigated by scanning electron microscopy (SEM) and
transmission electron microscopy (TEM) using a JEOL 5900 LV,
checking the high ordering of the hexagonal arrays and the large
aspect ratio. \cite{ELP+08} The chemical characterization of the
nanowires was made by means of energy dispersive analysis of
X-rays (EDAX), obtaining a clear peak corresponding to the
emission spectrum of Ni. \cite{ELP+08} The magnetic measurements
were performed by a vibrating sample magnetometer (VSM) with the
applied field at different angles. The measurements were made at
ambient temperature. Figure 1 illustrates the hysteresis curves
for two samples with diameter $d=$ 50 nm, lattice parameter $D=$
100 nm, and lengths \textit{(a)} $L=$ 4 $\mu $m and \textit{(b)}
12 $\mu $m, measured with the external field along and
perpendicular to the axis of the wires. In this figure it is
clearly seen that the easy axis of the nanowires is along the wire
axis.

\begin{figure}[h]
\begin{center}
\includegraphics[width=8cm]{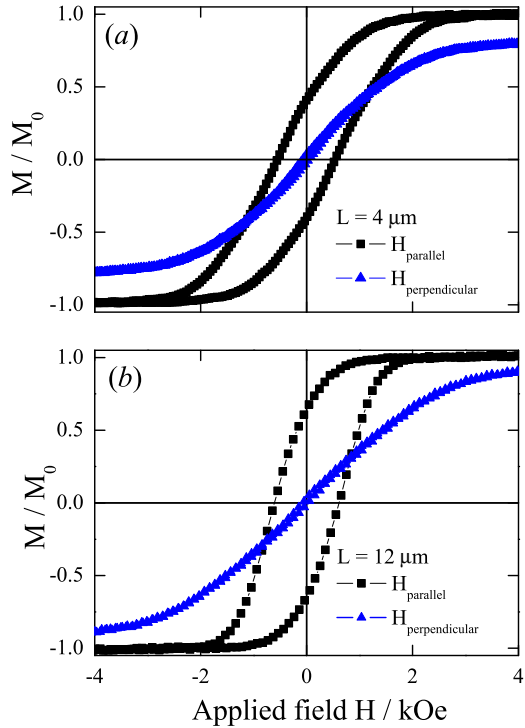}
\end{center}
\caption{Hysteresis curves at external field parallel and
perpendicular to the nanowire axis, for two arrays with (a) $L=$ 4
$\protect\mu $m and (b) 12 $\protect\mu $m, $d=$ 50 nm, and $D=$
100 nm.}
\end{figure}

In a series of Ni nanowire arrays of varying lengths (all other geometric
parameters being kept constant) investigated by VSM magnetometry, we
observed a significant dependence of the remanence (see Fig. 2) and
coercivity (see Fig. 3) on the angle at which the external magnetic field is
applied. In particular, the coercive field $H_{c}$ can be tuned between 100
and 720 Oe approximately by properly adjusting $\theta $. The 12 $\mu $m nanowires present a sharp decrease of the coercivity and remanence near $90%
%TCIMACRO{\U{b0}}%
%BeginExpansion
{{}^\circ}%
%EndExpansion
$, while the 4 $\mu $m sample has a broader curve.

\begin{figure}[h]
\begin{center}
\includegraphics[width=8cm]{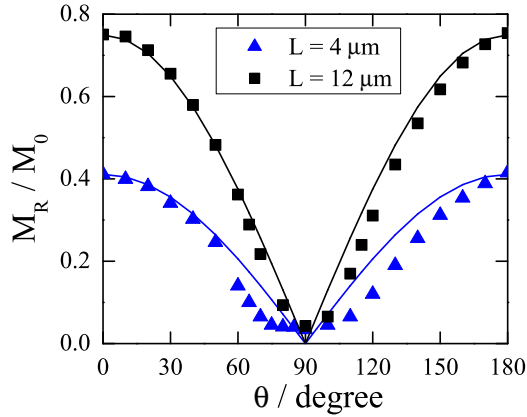}
\end{center}
\caption{Experimental results for the reduced remanence,
$M_{R}\left( \protect\theta \right) /M_{0}$ (where $M_{0}$ is the
saturation magnetization) as a function of the angle
$\protect\theta $ (dots), and analytical
expression given by $\left( M_{R}/M_{0}\right) \left\vert \cos \protect%
\theta \right\vert $, using the experimental values of the
intrinsic remanence, $M_{R}$, obtained from the hysteresis loops
of Fig. 1 (solid line).}
\end{figure}

\begin{figure}[h]
\begin{center}
\includegraphics[width=8cm]{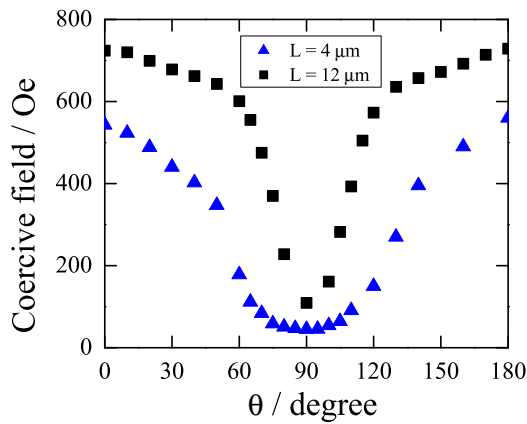}
\end{center}
\caption{Experimental data for the angular dependence of
coercivity in Ni
nanowire arrays, with length $L=$ 4 $\protect\mu $m and 12 $\protect\mu $m, $%
d=$ 50 nm, and $D=$ 100 nm.}
\end{figure}

\section{Analytical model}

In order to analyze our results for the angular dependence, we
made calculations which lead us to obtain analytical expressions
for both coercivity and remanence.

\subsection{Angular dependence of the remanence}

In an array the magnetization inside each nanowire is oriented
parallel to the wire's axis (easy axis), due to the strong shape
anisotropy. When the applied field is reduced to zero, at
remanence, each wire presents its magnetization along the axis,
but it is measured at an angle $\theta $ with respect to the easy
axis. When the field is reduced to zero, and due to the strong
shape anisotropy, the remanence rapidly relaxes and points along
the axis. Then, one can approximate the remanence of the wire by
the projection of the magnetization on the direction of
measurement, i. e., $M_{R}\left(
\theta \right) =M_{R}\left\vert \cos \theta \right\vert $, with $%
M_{R}=M_{R}\left( \theta =0\right) $ the remanence measured at
$\theta =0$. The solid line in Fig. 2 illustrates this expression
for $M_{R}\left( \theta\right)/M_{0} $ together with our
experimental results, showing a very good agreement, particularly
for $L=12$ $\mu m$, for which the shape anisotropy is stronger.
Then the angular dependence of the remanence can be described by
$M_{R}\left( \theta \right) =M_{R}\left\vert \cos \theta
\right\vert $ for nanowires with a large aspect ratio, i. e., with
their easy axis along the wire axis. For nanowires in which the
easy axis is perpendicular to the wire axis, the angular behavior
of the remanence can be modeled as $M_{R}\left\vert \sin \theta
\right\vert $. Note that this expression defines only the angular
behavior of the remanence, and does not determine the intrinsic
remanence, $M_{R}$, which has to be obtained by considering the
particular reversal mode, according to the geometry of the wire.

\subsection{Angular dependence of the coercivity}

In order to study the angular dependence of the reversal modes we have
developed analytical calculations that lead us to obtain the coercive field $%
H_{c}^{k}$ for each of the reversal mechanisms, $k=$ C, T, and V. The
angular dependence of the nucleation for a coherent magnetization reversal
was calculated by Stoner-Wohlfarth \cite{SW48} and gives%
\begin{equation*}
\frac{H_{n}^{C}\left( \theta \right) }{M_{0}}=-\frac{1-3N_{z}\left( L\right)
}{2}\frac{\sqrt{1-t^{2}+t^{4}}}{1+t^{2}}\ ,
\end{equation*}%
where $t=\tan ^{\frac{1}{3}}\left( \theta \right) $ and $M_{0}$ is the
saturation magnetization. The demagnetizing factor of a wire along the $z$
axis has been previously obtained \cite{BTZ+04, LEA+05} and is given by $%
N_{z}\left( l\right) =1-F_{21}\left[ \frac{4R^{2}}{l^{2}}\right] +\frac{8R}{%
3\pi l}$, where $F_{21}\left[ x\right] =F_{21}\left[ -1/2,1/2,2,-x\right] $
is a hypergeometric function.

For the T mode, the angular dependence of the coercivity can be studied by
an adapted Stoner-Wohlfarth model \cite{EDL+07, AEA+08} in which the width
of the domain wall, $w_{T}$, is used as the length of the region undergoing
coherent rotation. Starting from the equations presented by Landeros \textit{%
et al} \cite{LAE+07} we can calculate the width of the domain wall
for the transverse mode as a function of the wire geometry. The
domain wall width increases with the wire's radius, and is about
50 nm larger than the wire's radii. Following this approach,
\begin{equation*}
\frac{H_{n}^{T}\left( \theta \right) }{M_{0}}=-\frac{1-3N_{z}\left(
w_{T}\right) }{2}\frac{\sqrt{1-t^{2}+t^{4}}}{1+t^{2}}\ .
\end{equation*}%
It is important to mention that the expressions for $H_{n}^{T}\left( \theta
\right) $ and $H_{n}^{C}\left( \theta \right) $ differ only by the length.
In a coherent reversal, $L$ represents the total length of the wire, and
then the coercivity varies with the length. However, in a transverse mode,
and because $w_{T}$ is almost independent of the length of the wire, as
shown by Landeros \textit{et al} \cite{LAE+07}, the coercivity is also
independent of the length. Then, when $L>w_{T}$ ($L<w_{T}$), the $T$ ($C$)
mode will always exhibit a lower coercivity, independent of $\theta$.

As shown in the Stoner-Wohlfarth model \cite{SW48}, the nucleation
field does not represent the coercivity in all cases. However,
from the discussion presented on p. 21 in reference [19], the
coercivity for coherent and
transverse reversal modes can be written as%
\begin{equation*}
H_{c}^{C\text{(}T\text{)}}=\left\{
\begin{array}{c}
\left\vert H_{n}^{C\text{(}T\text{)}}\right\vert \quad 0\leq \theta \leq \pi
/4 \\
2\left\vert H_{n}^{C\text{(}T\text{)}}\left( \theta =\pi /4\right)
\right\vert -\left\vert H_{n}^{C\text{(}T\text{)}}\right\vert \text{ \ }\pi
/4\leq \theta \leq \pi /2%
\end{array}%
\right.
\end{equation*}%
In very short wires ($L\approx w_{T}$) the energy cost involved in the
creation of a domain wall gives rise to a coherent reversal.

The curling mode was proposed by Frei \textit{et al} \cite{FST57}
and has been used to investigate magnetic switching in films
\cite{IHS+97} and particles with different geometries, like
spheres \cite{FST57}, prolate ellipsoids \cite{FST57, ST63}, and
cylinders \cite{IS89}. However, for simplicity, expressions for
the nucleation field obtained using infinite cylinders are used in
all cases. The angular dependence of the curling
nucleation field in a finite prolate spheroid was obtained by Aharoni\cite%
{Aharoni97}.
\begin{equation*}
\frac{H_{n}^{V}}{M_{0}}=\frac{\left( N_{z}-\frac{q^{2}L_{x}^{2}}{R^{2}}%
\right) \left( N_{x}-\frac{q^{2}L_{x}^{2}}{R^{2}}\right) }{\sqrt{\left(
N_{z}-\frac{q^{2}L_{x}^{2}}{R^{2}}\right) ^{2}\sin ^{2}\theta _{0}+\left(
N_{x}-\frac{q^{2}L_{x}^{2}}{R^{2}}\right) ^{2}\cos ^{2}\theta _{0}}}\ .
\end{equation*}%
The demagnetizing factor of a wire along the $x$ axis has been
obtained previously \cite{BTZ+04, LEA+05} and is given by $N_{x}\left( l\right) =\frac{1%
}{2}F_{21}\left[ \frac{4R^{2}}{l^{2}}\right] -\frac{4R}{3\pi l}$. For a
cylindrical geometry, Shtrikman \textit{et al} \cite{ST63} have obtained $%
q^{2}=1.08\pi $. Here $J_{p}\left( z\right) $ and $Y_{p}\left( z\right) $
are Bessel functions of the first and second kind, respectively, and $L_{x}=%
\sqrt{2A/\mu _{0}M_{s}^{2}}$\ is the exchange length. As pointed
out by Aharoni \cite{Aharoni97}, for a prolate spheroid with
$\theta =0$, a jump of the magnetization at or near the curling
nucleation field occurs. Therefore, the coercivity is quite close
to the absolute value of the nucleation field. Then, we assumed
here that $-H_{n}^{V}$ is a good approximation to the coercivity,
$H_{c}^{V}$, when the reversal occurs in the V mode, as in other
studies\cite{ST63, I91}.

\begin{figure}[h]
\begin{center}
\includegraphics[width=8cm]{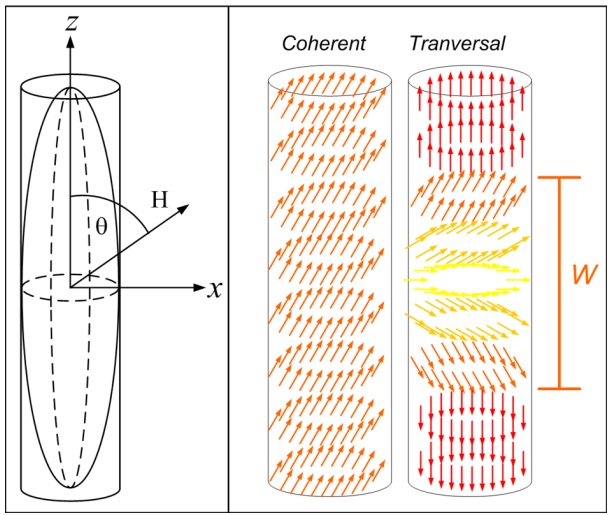}
\end{center}
\caption{Magnetization reversal modes in nanowires by coherent
rotation mode and transverse reversal mode.}
\end{figure}

\section{Results and discussion}

With the above expressions we can study the angular dependence of
the coercivity for our Ni nanowire array. In our model the system
reverses its magnetization by whichever mode opens an
energetically accessible route first, that is, by the mode that
exhibits the lowest coercivity. However, for highly dynamic cases,
the path of lowest coercivity might not be accessible due to
precession effects. By evaluating the coercivity for the different
modes described in Section III we found the one which drives the
reversal for each $\theta $. Figure 5 illustrates our results for
an isolated Ni nanowire with $L=12$ $\mu $m. The dotted line
represents the coercive field when the wire reverses its
magnetization by a coherent rotation; the dashed line is the
coercive field when the wire reverses its magnetization by the
propagation of a vortex domain wall, and the solid line represents
the coercive field when the wire reverses its magnetization by the
propagation of a transverse domain wall.

\begin{figure}[h]
\begin{center}
\includegraphics[width=8cm]{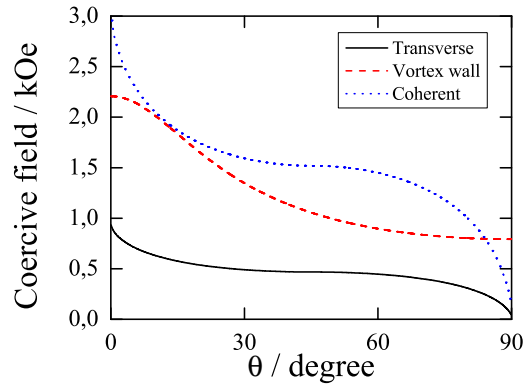}
\end{center}
\caption{Figure 5: Coercivity for different angles between the applied
magnetic field
and the wire axis (from 0%
%TCIMACRO{\U{ba} }%
%BeginExpansion
${{}^o}$
%EndExpansion
to 90%
%TCIMACRO{\U{ba}}%
%BeginExpansion
${{}^o}$%
%EndExpansion
).}
\end{figure}

As depicted in Fig. 5, in a 50 nm diameter Ni nanowire the T mode is
preferred for almost every $\theta$ because it exhibits a lower coercivity.
However, in the limit $\theta \sim 90$, the coercivity for the T and C modes
are very close.

Our results for Ni nanowires are combined in Fig. 6. The solid
line represents the coercivity obtained by means of analytical
calculations for $L=12$ $\mu $m. However, for almost every $\theta
$, differences with the results for $L=4$ $\mu $m are less than 3
$\%$, then at this scale, the line depicts results for both
lengths. Squares and
triangles represent the experimental values of the coercivity for $L=12$ $%
\mu $m and $L=4$ $\mu $m, respectively. As shown in Fig. 5, the
coercivity obtained from the Stoner-Wohlfarth model (coherent
rotation) is two to three times higher than the one obtained
assuming a transverse domain wall, for almost all $\theta$, and is
much larger than the experimental results. Then, although the
shape of the angular dependence of the coercivity is the same for
coherent rotation and the transverse reversal mode, the values are
significantly different, showing that a transverse mode drives the
reversal for each $\theta$.

\begin{figure}[h]
\begin{center}
\includegraphics[width=8cm]{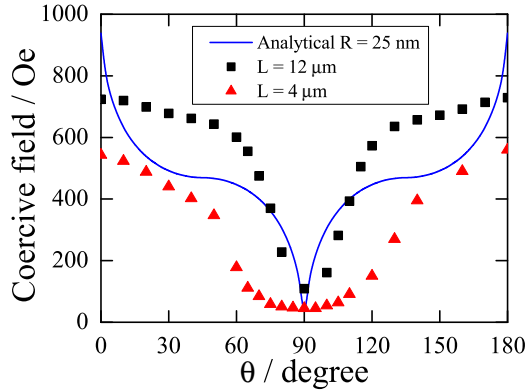}
\end{center}
\caption{Angular dependence of the coercive field measured
experimentally and calculated analytically for Ni nanowires with
50 nm diameter.}
\end{figure}

However, the absolute values computed for the coercivity are still
different compared with the experimental data. We ascribe such
difference between calculations and experimental results to the
interaction of each wire with the stray fields produced by the
array. When the distances between nanowires are smaller or
comparable to their diameter, the magnetostatic interactions
between nanowires are stronger, and consequently have an important
influence on the magnetic properties of the array. The interaction
of each wire with the stray fields produced by the array strongly
influences the coercivity \cite{ABD+97, Hertel01, BAA+06}.
Therefore, the small discrepancy between experiments and model can
be regarded as the result of interactions within the wires in the
array and size distribution, which are not included in our model.

Finally, the results presented above may be generalized. Figure 7
shows the angular dependence of the coercivity for an isolated Ni
nanowire when it reverses its magnetization by the propagation of
a transverse domain wall. We have considered different radii in
order to investigate the behavior of the geometry of the wire with
the angular dependence of the coercivity. In the range of radii
considered, we find that an increase in the radius $R$ results in
a decrease in the coercive field.

\begin{figure}[h]
\begin{center}
\includegraphics[width=8cm]{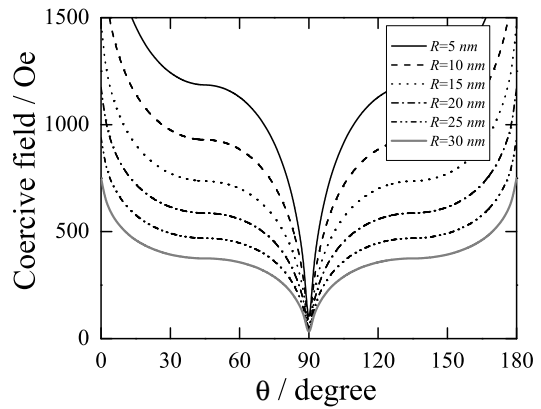}
\end{center}
\caption{Angular dependence of the coercivity in an isolated Ni
nanowire with different values of its radius. We have considered a
transverse reversal mode.}
\end{figure}

\section{Conclusions}

In conclusion, by means of theoretical studies and experimental
measurements we have investigated the angular dependence of the
magnetization in Ni nanowire arrays. As found from experiments,
the angular dependence in large aspect ratio wires can be
understood assuming that the relaxation process is very fast and
then the magnetization orients rapidly along the easy axes. We
have derived analytical expressions that allow us to obtain the
coercivity when the wire reverses its magnetization by means of a
coherent rotation, transverse reversal mode, and a vortex domain
wall. For the wires studied experimentally the magnetization is
driven by means of the nucleation and propagation of a transverse
wall. The stray field produced by the array, as a function of the
angle of the external field, cannot be determined analytically.
However, it is responsible for small differences between
experiments and calculations.

\section{Acknowledgments}

This work was supported by Financiamiento Basal para Centros Cientificos y
Tecnologicos de Excelencia, Millennium Science Nucleus \emph{Basic and
Applied Magnetism} P06-022F, Fondecyt Grants 11070010, 1080164, and 1080300,
and in part by the AFOSR under contract FA9550-07-1-0040. CONICYT Ph.D.
program and General Direction of Graduates at Universidad de Santiago de
Chile (DIGEGRA, USACH) are also acknowledged.

\end{document}